\documentclass[12pt]{article}
\usepackage{amsmath}
\usepackage{graphicx,psfrag,epsf}
\usepackage{enumerate}
\usepackage{url} 

\newcommand{\blind}{0}

\def\d{\mbox{d}}

\def\half{\hbox{$1\over2$}}

\def\N{\mbox{N}}

\def\P{\mbox{P}}
\def\d{{\rm d}}

\begin{document}



\def\spacingset#1{\renewcommand{\baselinestretch}%
{#1}\small\normalsize} \spacingset{1}

\if0\blind
{
\title{\bf A New Measure of Overlap: An Alternative to the $p$--value}
\author{Stephen G. Walker\hspace{.2cm}\\
Department of Mathematics \\
University of Texas at Austin, USA \\
e-mail: s.g.walker@math.utexas.edu}
\date{}
\maketitle       
} \fi   

\if1\blind
{
  \bigskip
  \bigskip
  \bigskip
  \begin{center}
    {\LARGE\bf A Note on the  Hypothesis Test for Comparing Normal Means}
\end{center}
  \medskip
} \fi




\bigskip
\begin{abstract}
In this paper we present a new measure for the overlap of two density functions which provides motivation and interpretation currently lacking with  benchmark measures based on the proportion of similar response, also known as the overlap coefficient. We use this new measure to present an alternative to the $p$--value as a guide to the choice of treatment in a comparative trial; where a current treatment and a new treatment are undergoing investigation. We show that it is possible to reject the null hypothesis; i.e. the new treatment is significantly different in response to the old treatment, while the proposed new summary for the same experiment indicates that as low as one in ten individuals subject to the new treatment behave differently to individuals on the old one. 
\end{abstract}

\vfill







\section{Introduction}  There has recently been a substantial amount of criticism of  the statistical hypothesis test, largely focusing on the {\em $p$--value}, though also on the notion of {\em significance}.
For example, see the recent discussion in the journal The American Statistician, Volume 73 (Supplementary Issue), headed by the Editorial; Wasserstein et al. (2019). Such discussions are by no means restricted to statistics journals; but are also prevalent in many scientific journals, with a good number appearing in medical journals, such as Cohen (2011).

To provide some background; the $p$--value is currently, though its future is in doubt, see Hasley (2019), the benchmark for reporting an outcome of an experiment. At a heuristic level, a small value indicates a significant outcome, i.e. to reject a null hypothesis, whereas a large value indicates a lack of significance. The cut--off mark, which has become the standard choice, is $0.05$. However, recently, this procedure has come under closer scrutiny and, despite much having been written on the subject, a consensus is far from apparent. 

Much of the  debate on $p$--values is generated through a lack of clear knowledge on what is being discussed, namely a rigorous definition of a $p$--value. Invariably, it is a heuristic definition based on ``seeing something larger than what was observed in a future experiment'' on which articles  predicate their discussion. However, such a heuristic fails even in a standard two--sided hypothesis test of a normal variance; see Gibbons and Pratt (1975). On the other hand, a rigorous definition of the $p$--value is provided by Lehmann  and Romano (2005); see also Kuffner  and Walker (2019).

With a rigorous definition of objects associated with a hypothesis test, there is nothing controversial about such a  test {\em per se}; see Savage (1972). The basic idea, well known, is to decide whether the observed statistic is compatible with the assumption of the hypothesis. The definition of ``compatible'' here amounts to setting a type I error, also known as the level of significance. This has typically been set at $0.05$. Such a specification amounts to acknowledging that from all true hypotheses tested, 1 in 20 will yield a significant and most likely non--reproducible result. It may be this lack of reproducibility which is really causing the unease among scientists; yet when decisions need to be made with imperfect information, mistakes can be made. That lack of reproducbility should immediately be classified as evidence of an irregular experiment, see for example Colquhoun (2018) and Nuzzo (2014), is unfortunate.

Once the hypothesis test framework has been determined to decide on a choice, concepts such as sufficient statistics, critical values, type I and II errors, power functions, become necessary parts of the decision process.
It is a theory well grounded, though most likely highly misunderstood among its many users. Nevertheless, the point behind the recent volume 73 of The American Statistician has been to think beyond the $p$--value and to consider alternative procedures for determining decisions related to outcomes of experiments; such as the comparison of a new treatment with an established treatment or with a placebo.  

The aim of this paper is to use the notion of {\sl overlap} as a means to compare treatment outcomes. Standard measures of overlap lack motivation in terms of interpretation. We resolve this problem by introducing a new measure of the overlap between two density functions;  the interpretation being the overlap measures the probability we can replace one outcome from one density with an outcome from the other density.  
Translating this idea to the experiment, it would be the probability of being able to replace one individual on one treatment with an individual on the other treatment. If the probability is 1, the treatments are deemed the same. If the probablity is 0, the treatments are as different as can be (though 0 and 1 would never be achieved in reality). Anything falling inbetween will provide a concrete interpretation as to the efficacy of the new treatment under investigation; i.e. what proportion of the population it benefits. This is the contribution of the paper. 

Our argument is that interpretability is everything. The use of statistics lacking clarity concerning outcomes, such as current measures of overlap, add to confusion. What is required is a statistic which has clear interpretation concerning outcome. 

For sake of clarity and ease of exposition, we discuss and demonstrate the new notion of overlap specifically to two normal density functions; though other distributions will be equally viable. If densities are not fully known, they are assumed to be estimable from data. One of the densities will the normal distribution under the null, current treatment, and the other normal density being the one of outcomes from the new treatment. Measures of overlap date back at least to Pearson (1895). In particular, Rom and Hwang (1996) employed the notion of overlap as a means of assessing similarity of treatments; the comparative trial becoming a common way of assessing the efficacy of two or more treatments. These authors used
the standard {\em Overlap Coefficient} for densities $p_0$ and $p_1$, defined as
\begin{equation}\label{overlap}
\mbox{OVL}(p_0,p_1)=\int \min\{p_0(x),p_1(x)\}\,d x;
\end{equation}
also known as the proportion of similar responses (PSR).
It is now quite a popular measure in medical data analysis; see, for example, Mizuno et al (2005), Lei and Olsen (2010), and Giacoletti and Heyse (2015). The measure (\ref{overlap}) is also used in ecology, measuring the so--called niche overlap; see, for example, Mason et al (2011). It also arises in economics, as a meaure of polarization between two groups; see Anderson (2004).
Apparently, in economics, it was first seen in a technical report by M. Weitzman in 1970 who was looking  at measures of overlap of income distributions. 

This said, Inman and Bradley (1989), state that the OVL,
has ``no major philosophical motivation''. For this reason, a formal use for detemining the choice of treatment from a comparative study will be elusive. With this as our cue, the main contribution of the present paper is to introduce a novel measure of overlap which has definitive and clear motivation. In fact, it represents the probability that a random outcome from one density can be replaced by a random outcome from the other density. Following PSR, we therefore label the new measure as the proportion of interchangeable responses (PIR). It is defined as
$$O_M(p_0,p_1)=\int \int \min\{p_0(x)p_1(y),p_0(y)p_1(x)\}\,d x\,d y,$$
which can be seen as a variant of OVL; i.e. $O_M(p_0,p_1)=\mbox{OVL}(p_a,p_b)$, where
$p_a(x,y)=p_0(x)p_1(y)$ and $p_b(x,y)=p_0(y)p_1(x)$. We believe $O_M$ to be  a new measure of overlap and provides a measure with better motivation and interpretation than that provided by PSR.
 
For the layout of the paper; in section 2 we set the background for the test. 
In section 3 we present the theory for how we compare outcomes from the two treatments, presenting a measure of quantification for a notion of overlap; i.e. the $O_M(p_0,p_1)$, and provide motivation for it. This intuitive measure of an outcome of an experiment can be severely at odds with the outcome of the hypothesis test. 
Section 4 presents an illustration and section 5 concludes with a brief discussion.

\section{Background on $p$--values }

To make the present paper concrete, we keep to a specific setting in which the use of the $p$--value is ubiquitous. Outcomes of a standard treatment $T_0$ among a population is known and represented by a normal distribution with known mean $\theta_0$ and known variance $\sigma^2$. Represent this via the density function
$p_0(x)$.

A new treatment $T_1$ is proposed and an experiment conducted on $n$ individuals from the population. The new treatment will present outcomes from a normal distribution with unknown mean $\theta$ and known variance $\sigma^2$. If indeed $\sigma$ is unknown, the argument which follows will apply, but with some more complicated technical details. 
The hypothesis of interest is
$$H_0:\theta=\theta_0\quad\mbox{vs}\quad H_1:\theta>\theta_0,$$ 
and, without loss of generality, we will take $\theta_0=0$. 
This is a standard set--up and the theory using $p$--values well documented. See, for example,  Cardinal (2016) and Pocock (2006).

Now $H_0$ will come under suspicion of being wrong if $\bar{X}$, the sufficient statistic for $\theta$, which is itself the sample mean of the observations from the individuals allocated to treatment $T_1$, is too large. Specifically, for a type I error of $\alpha$, the critical value is
$$c_{n,\alpha}=\frac{\sigma}{\sqrt{n}}\,\Phi^{-1}(1-\alpha),$$
where $\Phi$ is the standard normal distribution function.
That is, reject $H_0$ if $\bar{X}>c_{n,\alpha}$. 

The $p$--value corresponding to this experiment is given by the value of $\alpha$ which yields $\bar{X}=c_{n,\alpha}$; i.e.
$$p=1-\Phi\left(\frac{\bar{X}\,\sqrt{n}}{\sigma}\right).$$
Nothing new here; reject $H_0$ if $p<\alpha$.

The current debate about the $p$--value is whether it is appropriate as a summary outcome of an experiment, and  whether the benchmark $0.05$ is a good choice. A further complication is that the $p$--value is often misinterpreted and, even among statisticians, it has an ambiguous definition. See Panagiotakos (2008).


The underlying concern to the sciences is that if a mistake is made; i.e. the experiment is invoking the type I error, in which a null hypothesis is true yet the sufficient statistic lands in the critical region, which can happen with probability $\alpha$, then there is a lack of reproducibility. However, a level of lack of reproducibility is, as we have detailed, a factored aspect of the test. To elaborate further, suppose the 0.05 cut--off has been universally selected. Of all hypotheses which are true, one on twenty will yield a significant outcome. The up--shot is that one in twenty significant outcomes among experiments for which the null hypothesis is true, will not be reproducible. This seems a heavy price to pay. On the other hand, when decisions need to be made, mistakes can happen, inevitably.

While this concern has been well documented; the aim in the present paper is to provide a further criticism of the notion of the statistical test and to propose a solution.

\section{A new measure of overlap} 

We have assumed that the distribution of individual outcomes under the standard treatment $T_0$ is a normal distribution with mean 0 and variance $\sigma^2$, written as $p_0$. Under treatment $T_1$ the distribution of outcomes is normal with mean $\theta$ and variance $\sigma^2$, written as $p_1$.   Let $X_1$ denote a random outcome from $p_1$ and $X_0$ denote a random outcome from $p_0$. We want to quantify just exactly how $X_1$ compares to $X_0$. 

We do this in a universal way; so actually whatever the distribution of $X_0$ and $X_1$ are we can evaluate the measure of overlap we are about to describe. Let $N$ be a large number and let $(X_0^{(j)})_{j=1}^N$ be independent and identically distributed samples from $p_0$. For each $j$ we sample a $X_1^{(j)}$, and then according to some criterion either replace $X_0^{(j)}$ with $X_1^{(j)}$ or else leave it as it is. Hence, we end up with  a sample $(Z^{(j)})_{j=1}^N$ which will be distributed as $p_0$, as we shall see, and comprised of the ``accepted'' $(X_1^{(j)})$ and the left alone $(X_0^{(j)})$.

We define $q_N$ as the proportion of $(X_1^{(j)})$ in the sample $(Z^{(j)})$. We show that $q_N$ has a well defined limit as $N\to\infty$ which while difficult to compute directly, can always at least be estimated using Monte Carlo methods. 
We refer to $q$ as the overlap of $p_1$ into $p_0$. Effectively, the probability that a random outcome from $p_1$ can replace a random outcome from $p_0$; which we write as 
\begin{equation}\label{qdef}
q=\P(X_1\quad\mbox{r}\quad X_0).
\end{equation}
So if $q=1$ then $p_0$ and $p_1$ are identical; whereas if $q=0$ then the overlap of the support of the two densities is empty.

The $q$ is defined as
\begin{equation}\label{qval}
q=\int\int\min\{p_0(x)p_1(y),p_0(y)p_1(x)\}\,dx\,d y
\end{equation}
which, using the equality 
$|a-b|=a+b-2\min(a,b),$
can also be written as
$$
q=1-\half \int\int|p_0(x)\,p_1(y)-p_0(y)\,p_1(x)|\,d x\,d y,
$$
which is 0 if $p_0$ and $p_1$ have no common support points and is 1 if $p_0=p_1$. Note, as we would require, that this definition is symmetric in $p_0$ and $p_1$.
We now motivate this choice of $q$, which we also previously referred to as $O_M(p_0,p_1)$.

Consider the conditional density 
$$p(y|x)=\alpha(x,y)\,p_1(y)+(1-r(x))\,{\bf 1}(y=x),$$
where
\begin{equation}\label{alpham}
\alpha(x,y)=\min\left\{1,\frac{p_0(y)\,p_1(x)}{p_1(y)\,p_0(x)}\right\},
\end{equation}
and
$r(x)=\int \alpha(x,y)\,p_1(y)\,d y.$
Observers will recognize this as the transition density for a Metropolis--Hastings algorithm (Metropolis et al., 1953; Hastings, 1970; Tierney, 1994).
The point here is that if $X$ is distributed as $p_0$, and $Y$ is distributed as $p_1$, then 
$$Z=\left\{\begin{array}{ll}
Y & \mbox{with probability}\quad \alpha(X,Y) \\ \\
X & \mbox{with probability}\quad 1-\alpha(X,Y) 
\end{array}
\right.
$$
is distributed as $p_0$. 

Clearly of interest here is the probability we accept $Z$ as $Y$, since it replaces $X$, which is distributed as $p_0$, with $Y$ distributed as $p_1$, and recall $Z$ is also distributed as $p_0$. This probability is given by
$$q=\int r(x)\,p_0(x)\,\d x=\int\int \alpha(x,y)\,p_1(y)\,p_0(x)\,d x\,d y,$$
which is precisely (\ref{qval}). 
So, in short, the $q$ is the probability an outcome from $p_1$ replaces an outcome from $p_0$. Hence, if $q$ is large, the two treatments are not too different in that an outcome from an individual under $T_1$ can with high probability replace an outcome from an individual under $T_0$.
On the other hand, if $q$ is small, then the probability of this replacement is small indicating the outcomes from the two treatments are different. 

\section{Illustration}

Let us consider a specific case; taking $\sigma=1$, $n=100$, and $\alpha=0.05$. Under these settings the null hypothesis $H_0:\theta=0$ vs $H_1:\theta>0$ is rejected when
$$\bar{X}>\frac{1}{10}\,\Phi^{-1}(0.95)=0.164.$$
So now let us put $\theta=0.164$ and compute the value of $q$ which indicates just how differently $X_1$ under treatment $T_1$ behaves compared to $X_0$ under treatment $T_0$. 

We have $q=\mbox{P}(X_1\quad\mbox{r}\quad X_0)$ given by
$$q(\theta)=\int\int \min\{\phi(x)\,\phi(y-\theta),\,\phi(y)\,\phi(x-\theta)\}\,d x\,d y,$$
where $\phi(x-\theta)$ is the normal density function with mean $\theta$ and variance 1.
In general with any choice of $p_0$ and $p_1$ it would be difficult to compute $q$ directly,
and so it would need to be computed using Monte Carlo methods. 
However, for the normal case here it is computable directly and
$$q(\theta)=2\left(1-\Phi\left(\frac{\theta}{\sqrt{2}}\right)\right).$$
With $\theta=0.164$, we get $q= 0.91$.  

This is an interesting result. It means that while $H_0$ is rejected at the $5$\% level of significance for this value of $\theta=0.164$, indicating $T_1$ is preferred to $T_0$; on the other hand, also true is that only 1 in 10 individuals behave differently under $T_1$ than they do under $T_0$. On this basis we question the relevance of the outcome of the hypothesis test and indeed whether the test is in itself a reasonable course of action to take.
As espoused by Dom and Hwang (1996), a measure of similarity would seem more insightful and informative.


\begin{center}
\begin{figure}[!htbp]
\begin{center}
\includegraphics[width=14cm,height=10cm]{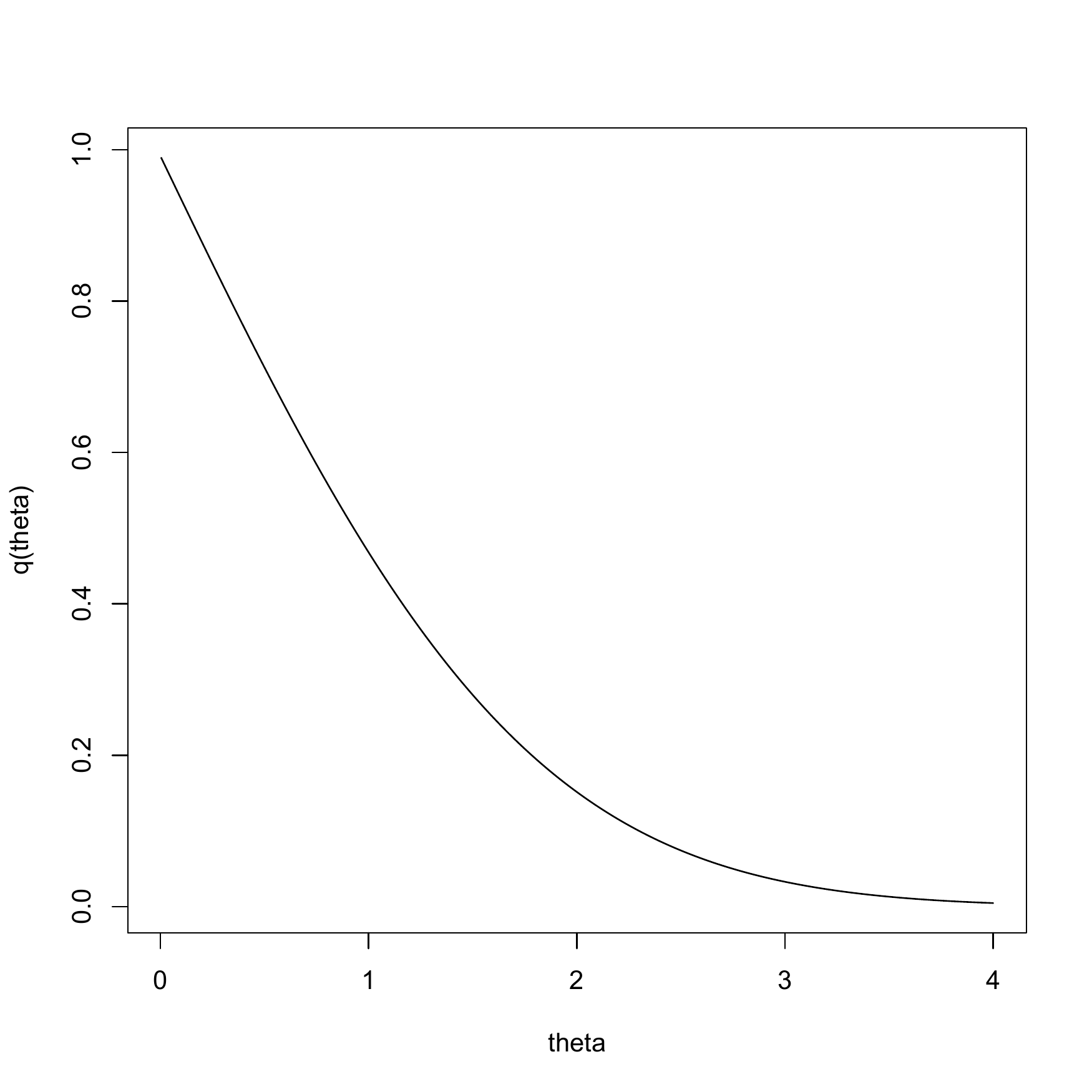}
\caption{Plot of $q(\theta)$ for varying $\theta$ in normal illustration}
\label{fig1}
\end{center}
\end{figure}
\end{center} 

Here we plot the probabilty $q(\theta)$ as a function of $\theta$ for $\theta>0$, and is provided in Fig.~\ref{fig1}.
One can see that the probability an individual under treatment $T_1$ replacing an individual under treatment $T_0$ dips below $\half$ for $\theta\approx 1$; a value substantially larger than 0.164.

In practice, the experiment would yield $\bar{X}$, the estimate of the mean $\theta$ from the observations. As a consequence, if there is a need to avoid levels of significance and $p$--values, one can favor $T_1$ if $\bar{X}>1$, for example. Under this scenario there is no Type I error and no notion of a probability of making an error. An evaluation is being made of treatment $T_1$ and how it relates in terms of performance compared to $T_0$. The question becomes whether this performance is sufficiently different to merit changing treatment.

In general, if it is determined that $\P(X_1\quad\mbox{r}\quad X_0)<q_0$ for some specified $q_0$, 
then one would accept the new treatment if
$q(\theta)<q_0$; in the example, this would be $\theta>\sqrt{2}\Phi^{-1}(1-q_0/2)$. With a Bayesian or bootstrap approach, a probability for this event can be calculated. Alternatively, one could test the hypothesis
$H_0:\,\theta>q^{-1}(q_0).$

An interesting note to add here is that to recover the hypothesis $H_0:\theta>0$, to determine whether the new treatment is better or not, would be tantamount to selecting $q_0=1$, which makes no sense. The point being that the hypothesis $H_0:\theta>0$ can be rejected even with a $\theta$ being marginally larger than 0 which equates to essentially no benefit. Whereas by asking that for some $q_0$ it is demonstrated that $\theta>q^{-1}(q_0)$ not only has a benefit been suggested but also one of sufficient magnitude.

\section{Summary and discussion}

The hypothesis test, with associated summaries such as the $p$--value and level of significance, has been shown to be deficient  as a means to determing whether a new treatment is {\em suitably} different to an existing one. It has been argued, by many authors, that decisions based on $p$--values and statistical significance have many shortcomings. 

A further shortcoming of the relevance of the $p$--value and level of significance has been presented in this paper. Indeed, we argue that the overlap of outcomes, as measured by a probability $q$, under the two treatments is a highly relevant statistic of an experiment, shedding light by computing the probability an individual under the new treatment can replace an individual on the old treatment.  This is given by the $q$ in (\ref{qval}), where $p_0$ is the density of outcomes under $T_0$ and $p_1$ the (estimated) density of outcomes under $T_1$. That is, if $\widehat{\theta}$ is the sample estimate of $\theta$, we compute the probability of a random outcome from $p_{\widehat{\theta}}$ being a replacement for a random outcome from $p_0$ as
$$q=\int\int\min\{p_0(x)\,p_{\widehat{\theta}}(y),p_0(y)\,p_{\widehat{\theta}}(x)\}\,d x\,d y.$$ 
This can be close to 1, indicating no substantial difference between the density functions, whereas the $\widehat{\theta}$ lies in the critical region for rejecting $H_0$. 

A criticism of this $q$ as it stands is that it fails to take the uncertainty in the data into account, relying as it does on $\widehat{\theta}$. The bootstrap (Efron, 1979; Efron, 2012) would be one way to take the uncertainty of $\widehat{\theta}$ into consideration. This would involve, for the parametric bootstrap, the most suitable bootstrap under the circumstances, sampling a new data set $(\widetilde{X}_1,\ldots,\widetilde{X}_n)$ independently and identically distributed from $p_{\widehat{\theta}}(.)$ and computing the maximum likelihood estimator $\widetilde{\theta}$ from this sample. Each $\widetilde{\theta}$ would provide a 
$$\widetilde{q}=\int\int\min\{p_0(x)\,p_{\widetilde{\theta}}(y),p_0(y)\,p_{\widetilde{\theta}}(x)\}\,d x\,d y.$$
A collection of the $(\widetilde{q})$, an arbitrarily large number, can be used to account now for the uncertainty in $\widetilde{\theta}$.


It is worth taking the value of $q$ as a decision summary for the experiment alone. An approach considered in 
Resier and Faraggi (1999), who tested the equivalence of two populations using PSR.  For example, if more than half of individuals under the new treatment behave as though they were on the old treatment, the new treatment might be judged not good enough. Nevertheless, in the context of the problem for which the treatments are being used, it might be that even a small proportion of individuals behaving sufficiently differently makes the switch from old to new justifiable. 

Comments on an earlier version of this paper suggest that a large scale simulation study is in order. This misses the point which is that a motivated statistic is required. One may or may not think in any given experiment that the specific motivation is appropriate. But whichever opinion is presented, a table of numbers providing measures of overlap which have no interpretation only adds to a sense of statistics being used out of place.

\section*{Appendix} We present a number of pieces of further information about the new measure of overlap. In Appendix A we consider the distance associated with the new measure of overlap and connect it to the Hellinger distance. In Appendix B  we consider yet an alternative overlap measure which can be seen as being motivated by an alternative to the Metropolis--Hastings algorithm; the Barker algorithm. In Appendix C we provide a bound between $O_M$ and $OVL$ and in Appendix D we connect up the new overlap measure with the Youden index. Finally, in Appendix E  we use the overlap measure to provide a new measure of overlap between two finite sets.

\subsection*{Appendix A} Here we discuss more formally the $q=\mbox{P}(X_1\quad \mbox{r} \quad X_0)$ as defined by 
$$q=\int\int \min\{p_0(x)\,p_1(y),p_0(y)\,p_1(x)\}\,d x\,d y.$$
First note the necessary property that $\mbox{P}(X_1\quad \mbox{r} \quad X_0)=\mbox{P}(X_0\quad \mbox{r} \quad X_1)$; since if $X_0$ can replace $X_1$ with probability $q$, so with the same probability $X_0$ can replace $X_1$. 
Also note that
$$d(p_0,p_1)=1-\int\int \min\{p_0(x)\,p_1(y),p_0(y)\,p_1(x)\}\,d x\,d y$$
is a distance between density functions $p_0$ and $p_1$.

So the idea is that while $\theta_0$ and $\widehat{\theta}$ are far enough apart for the null hypothesis to be rejected, the distance between $p_{\theta_0}$ and $p_{\widehat{\theta}}$ is small. However, it is imperative the distance used to compare $p_{\theta_0}$ and $p_{\widehat{\theta}}$ is interpretable and is calibrated in terms of a relevant unit relating to treatment outcomes. This is the idea being $q=\mbox{P}(X_1\quad \mbox{r} \quad X_0)$.


We can also show that $d(p_0,p_1)$ is equivalent to the Hellinger distance
$$d_H(p_0,p_1)=\left(\int \left(\sqrt{p_0(x)}-\sqrt{p_1(x)}\right)^2\,d x\right)^{1/2}$$
and so
$$\int \sqrt{p_0(x)\,p_1(x)}\,d x=1-\half d_H^2(p_0,p_1).$$
Recall, setting $p(x,y)=p_0(x)\,p_1(y)$, 
$$d(p_0,p_1)=\half \int\int |p(x,y)-p(y,x)|\,d x\,d y=1-\int\int\min\{p(x,y),p(y,x)\}\,d x\,d y.$$
Using 
$$1-a/b=(1-\sqrt{a/b})(1+\sqrt{a/b}) \quad\mbox{and}\quad \min\{a,b\}\leq\sqrt{ab},$$
we can show that
$$d_H^2(p_0,p_1)\left(1-d_H^2(p_0,p_1)/4\right)\leq d(p_0,p_1)\leq 4d_H^2(p_0,p_1).$$

\subsection*{Appendix B} An alternative to the Metropolis--Hastings algorithm would be the Barker algorithm (Barker, 1965). Here instead of the $\alpha$ appearing in (\ref{alpham}), we have
$$\alpha(x,y)=\frac{p_0(y)p_1(x)}{p_0(y)p_1(x)+p_0(x)p_1(y)},$$
and not that
$$\alpha(x,y)\,p_0(x)p_1(y)=\alpha(y,x)\,p_0(y)p_1(x).$$
The measure of overlap in this case is
$$O_B(p_0,p_1)=2\int\int\frac{p_0(y)p_1(x)\,p_0(x)p_1(y)}{p_0(y)p_1(x)+p_0(x)p_1(y)}\,d x\,d y,$$
the specific form ensuring that the measure is 1 when $p_0\equiv p_1$ and $O_B$ is upper bounded by 1. It is interesting to note that this measure is connected to the {\sl Crossmatch} statistic (Rosenbaum, 1995). The crossmatch statistic is for testing the equality of two distributions represented by two independent samples $X_{1:n}$ and $Y_{1:n}$. The statistic is the number of pairings between the two samples which minimizes a sum of distances, and the limit of the statistic, as $n\to \infty$, is given by
$$O_C(p_0,p_1)=2\int \frac{p_0(x)p_1(x)}{p_0(x)+p_1(x)}\,d x,$$ 
where $p_0$ is the density of the $X$'s and $p_1$ the density of  $Y$'s. So just as we obtained $O_M$ from OVL, so we obtain $O_B$ from $O_C$; i.e. $O_B(p_0,p_1)=O_C(p_0(x)p_1(y),p_0(y)p_1(x))$. 

Now $O_B$ does not need to be estimated by first estimating the appropriate density functions, which is needed for the overlap coefficient. Such estimation of the densities can be done using kernel smoothing, as in Stine and Heyse (2001). Consider data $X_{1:n}$ and $Y_{1:n}$ from $p_0$ and $p_1$, respectively. To estimate $O_B$ directly from the samples, consider the two bivariate samples
$$A=\left(\begin{array}{lll}
X_1 & \cdots & X_{n/2} \\
Y_1 & \cdots & Y_{n/2}
\end{array}\right)\quad\mbox{and}\quad B=\left(\begin{array}{lll}
Y_{n/2+1} & \cdots & Y_n \\
X_{n/2+1} & \cdots & X_n
\end{array}\right)$$
and construct the $n\times n$ symmetric matrix $M$ with elements
$$M(i,j)=d_E(A_i,A_j),\quad M(i,n/2+j)=d_E(A_j,B_j),\quad M(n/2+i,n/2+j)=d_E(B_i,B_j),$$
for $1\leq i,j\leq n/2$, where $d_E$ denotes the usual Euclidean distance. 
Then evaluate the idempotent permutation $\sigma$ on $\{1,\ldots,n\}$ which minimizes
$l(\sigma)=\sum_{i=1:n} M(i,\sigma(i)).$
The relevant statistic which converges to $O_B$ is $\min\{1,4n_c/n\}$, where
$n_c=\#\{i:\sigma(i)>n/2\}$. See Arias--Castro and Pelletier (2016).

\subsection*{Appendix C} Here we find a bound between $O_M(f,g)$ and $OVL(f,g)$. Now $\min\{a,b\}\leq \sqrt{ab}$ so
$$O_M(f,g)\leq \left(\int\sqrt{f(x)\,g(x)}\,d x\right)^2.$$
Also, $|a-b|=a|1-\sqrt{b/a}|(1+\sqrt{b/a})$, and so using Jensen inequality
$$\int |f-g|\leq 2\sqrt{1-\left(\int\sqrt{fg}\right)^2 }.$$
Hence,
$$O_M(f,g)\leq \left( 1-\left(\half \int |f-g|\right)^2 \right)^2$$
and using $|a-b|=a+b-2\min(a,b)$ we recover
$$O_M(f,g)\leq \left(1-\left(1-OVL(f,g)\right)^2\right)^2.$$

\subsection*{Appendix D} In Samawi et al (2017) a relation between the Youden index (Youden, 1950) and the OVL overlap measure was explained. The Youden index for densities $f$ and $g$, with corresponding probability measures/distribution functions $F$ and $G$, respectively, and with cut--off mark $c$, is given by
$$J(c)=F(A(c))+\bar{G}(A(c)),$$
where $\bar{G}\equiv 1-G$ and $A(c)=(-\infty,c)$. One aim here is to estimate $c$ by maximizing $J(c)$; see, for example, Fluss et al (2005).

Now $OVL(f,g)$ can be written as
$$OVL(f,g)=F(A)+\bar{G}(A)$$
where $A=\{x:f(x)/g(x)\leq 1\}$. In fact, if we considered
$J(A)=F(A)+\bar{G}(A)$
and attempted to maximize this over all sets $A$, the solution would indeed be the set $\{x:f(x)/g(x)\leq 1\}$.

The $O_M(f,g)$ measure is more general again; it can be written as
$$O_M(f,g)=\int F(A(y))\,g(y)\, dy+\int \bar{G}(A(y))\,f(y)\,d y,$$
where
$$A(y)=\{x:f(x)/g(x)\leq f(y)/g(y)\}.$$
In this context, $O_M(f,g)$ can be seen as a natural extension of the Youden index and OVL measure.

\subsection*{Appendix E} Consider two finite sets $A$ and $B$. The overlap coefficient is a measure of overlap between the two sets and given by
$$O(A,B)=\frac{|S|}{\min\{|A|,|B|\}}$$
where $S=A\cap B$ and $|\cdot|$ denotes the size of the set. An alternative measure is the Jaccard index
$$J(A,B)=\frac{|S|}{|A|+|B|-|S|}.$$
Overlap or similarity measures are routinely used in medical imaging; see for example, Almodovar--Rivera and Maitra (2019).
 
We can use $O_M$ here by creating two discrete probability mass functions $p_A$ and $p_B$ by putting mass $1/|A|$ on each element of $A$ and mass $1/|B|$ on each element of $B$, respectively. Then, define the new measure of overlap for two finite sets $A$ and $B$ as
$$O_M(A,B)=\sqrt{O_M(p_A,p_B)}=\sqrt{\sum_{i=1}^{|A|}\sum_{j=1}^{|B|}\frac{\min\{p_{iA}p_{jB},p_{iB}p_{jA}\}}{|A||B|}}=\frac{|S|}{\sqrt{|A|\,|B|}}.$$
It is now interesting to note that $O_M$ sits precisely in between $O$ and $J$; i.e.
$$J(A,B)\leq O_M(A,B)\leq O(A,B).$$
This follows since for any integers $n,m$ with $k\leq\min\{n,m\}$,
$$k\leq \sqrt{nm}\leq n+m-k.$$

\section*{References}


\begin{description}

\item Almodovar--Rivera, I. and Maitra, R. (2019). FAST adaptive smoothing and thresholding for improved activation detection in low-signal fMRI, {\em IEEE Transactions on Medical Imaging}, 38, 2821--2828.

\item Anderson, G.J. (2004), Toward an empirical analysis of polarization, {\em Journal of Econometrics} 122, 1--26.

\item Barker, A.A. (1965), Monte Carlo calculations of the radial distribution functions for a proton--electron plasma, {\em Australian Journal of Physics}, 18, 119--133. 

\item Cardinal, L.J. (2016), Determining true difference between treatment groups,
{\em Journal of Community Hospital Internal Medicine  Perspectives}, 6, 10.3402/jchimp.v6.30284. 

\item Arias--Castro, E. and Pelletier, B. (2016), 
On the consistency of the crossmatch test, 
{\em Journal of Statistical Planning and Inference} 171, 184--190.

\item Cohen, H.W. (2011),
$P$--values: Use and misuse in medical literature,
{\em American Journal of Hypertension}, 24, 18--23.

\item Colquhoun, D. (2018),
Reproducibility of science: Fraud, impact factors and carelessness,
{\em Journal of Molecular and Cellular Cardiology}, 114, 364--368.

\item Efron, B. (1979), Bootstrap methods: another look at the jackknife, {\em Annals of Statistics}, 7, 1--26.

\item Efron, B. (2012), Bayesian inference and the parametric bootstrap, {\em Annals of Applied Statistics}, 4, 1971--1997.

\item Fluss, R., Faraggi, D. and Reiser, B. (2005), Estimation of the Youden index and its associated cutoff point, {\em Biometrical Journal} 4, 458--472.

\item Giacoletti, K.E.D. and Heyse, J. (2015), Using proportions of similar response to evaluate correlates of protection for vaccine efficacy, {\em Statistical Methods in Medical Research} 24, 273--286. 

\item Gibbons, J.D. and Pratt, J.W., (1975),
$P$--values: Interpretation and methodology,
{\em The American Statistician}, 29, 20--25.

\item Hasley, L.G. (2019),
The reign of the $p$--value is over: what alternative analyses could we employ to fill the power vacuum?,
{\em Biology Letters}, 15, 2018174.

\item Hastings, W.K. (1970), Monte Carlo sampling methods using Markov chains and their applications, {\em Biometrika}, 57, 97--109.

\item Inman, H.F. and Bradley, E.L. (1989). The overlapping coefficient as a measure of agreement between probability distributions and point estimation of the overlap of two normal densities. {\em Communications in Statistics -- Theory and Methods}, 18, 3851--3874. 

\item Kuffner, T. and Walker, S.G., (2019), Why are $p$--values controversial?,
{\em The American Statistician}, 73, 1--3.

\item Lehmann. E.L. and Romano, J.P. (2005),
{\em Testing Statistical Hypotheses}, Third Edition, Springer Texts in Statistics.

\item Lei, L. and Olsen, K. (2010), Evaluating statistical methods to establish clinical similarity of two biologics, {\em Journal of Biopharmaceutical Statistics} 20, 62--74. 

\item Mason, N.W.H., de Bello, F., Dolezal, J. and Leps, J. (2011), Niche overlap reveals the effects of competition, disturbance and contrasting assembly processes in experimental grassland communities, {\em Journal of Ecology} 99, 788--796.

\item Metropolis, N., Rosenbluth, A.W., Rosenbluth, M.N., and Teller, A.H. (1953), Equation of state calculations by fast computing machines, {\em Journal of Chemical Physics}, 21, 1087--1092.

\item Nuzzo, R. (2014), Scientific method: Statistical errors,
{\em Nature}, 506, 150--152.

\item Panagiotakos, D.B. (2008), The value of $p$--value in biomedical research,
{\em The Open Cariovascular Medicine Journal}, 2, 97--99.

\item Pearson, K. (1895), Contributions to the mathematical theory of evolution, II: skew variation in homogeneous matterial, {\em Philosophical Transactions of the Royal Society of London, Series A}, 186, 343--414. 

\item Pocock, S.J. (2006), The simplest statistical test: how to check for a difference between treatments,
{\em The British Medical Journal}, 332, 1256--1258.

\item Resier, B. and Faraggi, D. (1999), Confidence intervals for the overlapping coefficient: the normal equal variance case, {\em The Statistician} 48, 413--418.

\item Rom, D.M. and Hwang, E. (1996), Testing for individual and population equivalence based on the proportion of similar response, {\em Statistics in Medicine} 15, 1489--1505.

\item Rosenbaum, P. R. (2005), An exact distribution‐-free test comparing two multivariate distributions based on adjacency, {\em Journal of the Royal Statistical Society: Series B (Statistical Methodology)}, 67,  515--530.

\item Samawi, H.M., Yin, J., Rochani, H. and Panchal, V. (2017), Notes on the overlap measure as an alternative to the Youden index: How are they related? {\em Statistics in Medicine} 36, 4230--4240.

\item Savage, L.J. (1972). {\em The Foundations of Statistics},
Second Revised Edition, Dover Publications, NY.

\item Stine, R.A. and Heyse, J.F. (2001), Non--parametric estimates of overlap, {\em Statistics in Medicine} 20, 215--236.

\item Tanaka--Mizuno, S., Yamaguchi, T., Fukushima, Y., and Matsuyama, Y. (2005),  Overlap coefficient for assessing the similarity of pharmacokinetic data between ethnically different populations, {\em Clinical Trials} 2, 174--181.

\item Tierney, L. (1994), Markov chains for exploring posterior distributions, {\em The Annals of Statistics}, 22, 1701--1728.

\item Wasserstein, R.L., Schirm, A.L. and Lazar, N.A. (2019), Moving to a world beyond $p<0.05$,
{\em The American Statistician}, 73(Supp), 1--19. 

\item Youden, W.J. (1950), Index for rating diagnostic tests, {\em Cancer} 3, 32--35.

\end{description}

\end{document}